\documentstyle[12pt,fleqn]{article}
\topmargin - 0.8cm

\catcode`@=11
\@addtoreset{equation}{section}
\catcode`@=12
\mathchardef\SGamma="7100

\begin{document}
\title{\vskip-1.7cm \bf  Nonlocal action for long-distance modifications
of gravity theory}
\date{}
\author{A.O.Barvinsky$^{\dag}$}
\maketitle
\hspace{-8mm}
{\em Theory Department, Lebedev Physics
Institute, Leninsky Prospect 53, Moscow 117924, Russia}

\begin{abstract}
We construct the covariant nonlocal action for recently suggested
long-distance modifications of gravity theory motivated by the
cosmological constant and cosmological acceleration problems. This
construction is based on the special nonlocal form of the
Einstein-Hilbert action explicitly revealing the fact that this
action within the covariant curvature expansion begins with
curvature-squared terms. Nonlocal form factors in the action of
both quantum and brane-induced nature are briefly discussed. In
particular, it is emphasized that for certain class of quantum
initial value problems nonlocal nature of the Euclidean action
does not contradict the causality of effective equations of
motion.
\end{abstract}
$^{\dag}$e-mail: barvin@td.lpi.ru

\section{Introduction}
\hspace{\parindent} The purpose of this paper is to suggest the
class of nonlocal actions for covariantly consistent infrared
modifications of Einstein theory discussed in \cite{AHDDG}. The
modified equations of motion were suggested to have the form of
Einstein equations
    \begin{eqnarray}
    M_P^2\,\Big(1+{\cal F}(L^2\Box\,)\Big)
    \left(R_{\mu\nu}-\frac12
    g_{\mu\nu}R\right)=\frac12\,T_{\mu\nu}  \label{1.1}
    \end{eqnarray}
with "nonlocal" inverse gravitational constant or Planck mass,
$M_P^2(\Box)=M_P^2\,\big(1+{\cal F}(L^2\Box\,)\big)$, being some
function of the dimensionless combination of the covariant
d'Alem\-bertian $\Box=g^{\alpha\beta}\nabla_\alpha\nabla_\beta$
and the additional scale $L$ -- the length at which infrared
modification becomes important, $1/\sqrt{-\Box}\sim L$. If the
function of this dimensionless combination $z=L^2\Box$ satisfies
the conditions
    \begin{eqnarray}
    &&{\cal F}(z)\to 0,\,\,\,\,\,\,\,\,
    z\gg 1\nonumber\\
    &&{\cal F}(z)\to {\cal F}(0)\gg 1,
    \,\,\,\, z\to 0,                           \label{1.2}
    \end{eqnarray}
then the long-distance modification is inessential for processes
varying in spacetime faster than $1/L$ and is large for slower
phenomena at wavelengthes $\sim L$ and larger. This opens the
prospects for resolving the cosmological constant problem,
provided one identifies the scale $L$ with the horizon size of the
present Universe $L\sim 1/H_0\sim 10^{28} cm$. Indeed, equations
(\ref{1.2}) then interpolate between the Planck scale of the
gravitational coupling constant $G_P=16\pi/M^2_P$ for local matter
sources of size $\ll L$ and the long distance gravitational
constant $G_{LD}=16\pi/M^2_P(1+{\cal F}(0))\ll G_P$ with which the
sources nearly homogeneous at the horizon scale $L$ are
gravitating. Therefore, the vacuum energy ${\cal E}$,
$T_{\mu\nu}={\cal E}g_{\mu\nu}$, of TeV or even Planckian scale
(necessarily arising in all conceivable models with spontaneously
broken SUSY or in quantum gravity) will not generate a
catastrophically big spacetime curvature incompatible with the
tiny observable $H_0^2$. This mechanism is drastically different
from the old suggestions of supersymmetric cancellation of ${\cal
E}$ \cite{Weinberg}, because it relies on the fact that the nearly
homogeneous vacuum energy gravitates very little, rather than it
is itself very small\footnote{The idea of the scale dependent
Newton's constant was also suggested in \cite{ParSol} within
curved-brane models, though it was not explicitly formulated in
terms of a nonlocal form factor.}. It will generate the curvature
$H^2\sim G_{LD}{\cal E}\sim G_P\,{\cal E}/{\cal F}(0)$ which can
be very small due to large ${\cal F}(0)$.

Various aspects of this idea have been discussed in much detail in
\cite{AHDDG}. One formal difficulty with this construction was
particularly emphasized by the authors of \cite{AHDDG}. Point is
that for any nontrivial form factor ${\cal F}(L^2\Box\,)$ the left
hand side of (\ref{1.1}) does not satisfy the Bianchi identity
and, therefore, cannot be generated by generally covariant action.
The equations (\ref{1.1}) are generally covariant, but cannot be
represented as a metric variational derivative of the
diffeomorphism invariant action. Obviously, this makes the
situation unsatisfactory because of a missing off-shell
extension of the theory, problems with its quantization, etc.

In this paper we suggest to circumvent this problem by the
following simple observation. Point is that the infrared regime,
which is crucial for the resolution of the cosmological constant
problem, implies not only the long-wavelength but also the {\em
weak field} approximation. This means that the equation
(\ref{1.1}) is literally valid only as a first term of the
perturbation expansion in powers of the curvature. Therefore, the
left hand side of (\ref{1.1}) should be modified by higher than
linear terms in the curvature, and the modified nonlocal action
$S_{NL}[\,g\,]$ should be found from the variational equation
    \begin{eqnarray}
    \frac{\delta S_{NL}[\,g\,]}{\delta g_{\mu\nu}(x)}=
    M_P^2\,g^{1/2}\Big(1+{\cal F}(L^2\Box\,)\Big)
    \left(R^{\mu\nu}-\frac12
    g^{\mu\nu}R\right)+{\rm O}\,[\,R_{\mu\nu}^2\,]. \label{1.3}
    \end{eqnarray}
Flexibility in higher orders of the curvature allows one to
guarantee the integrability of this equation and to construct the
nonlocal action as a generally covariant (but nonlocal) curvature
expansion. Here we explicitly present this construction along the
lines of covariant curvature expansion developed in
\cite{beyond,CPTII,basis} a number of years ago. As a starting
point we consider a special nonlocal form of the Einstein-Hilbert
action revealing its basic property -- the absence of a linear in
metric perturbation part (on flat-space background), which is
apparently the classical analogue of a tadpole elimination
technique in non-SUSY string models \cite{tadpole}. Then we
introduce a needed long-distance modification by a simple
replacement of the nonlocal form factor in the curvature-squared
term of the obtained action. The paper is accomplished by a
discussion of the nature of nonlocalities in quantum-gravitational
and brane-induced models of \cite{GRS,DGP,DR}. In particular, the
fact that curvature expansion for the action begins with the
quadratic order is revisited from the viewpoint of the running
gravitational coupling constant. Finally, the issue of acausality
of nonlocal effective equations, raised in \cite{AHDDG}, is
reconsidered and a possible generalization to asymptotically
deSitter spacetimes is briefly discussed.

\section{Nonlocal form of the Einstein action}
\hspace{\parindent} For simplicity, we start with the Euclidean
(positive-signature) asymptotically-flat spacetime in $d$
dimensions. The action of Einstein theory
    \begin{eqnarray}
    S_E[\,g\,]=-M_P^2
    \int dx\,g^{1/2}\,R(\,g\,)
    -2\,M_P^2\int_\infty\!
    d^{d-1}\sigma\,\Big(g^{(d-1)}\Big)^{1/2}\,
    \Big(K-K_0\Big)                 \label{2.1}
    \end{eqnarray}
includes the bulk integral of the $d$-dimensional scalar curvature
and the surface integral over spacetime infinity, $|x|\to\infty$,
with induced metric $g^{d-1}$. The latter is usually called the
Gibbons-Hawking action which in the covariant form contains the
trace of the extrinsic curvature of the boundary $K$ (with the
subtraction of the flat space background $K_0$). This surface term
guarantees the consistency of the variational problem for this
action which yields as a metric variational derivative the
Einstein tensor
    \begin{eqnarray}
    \frac{\delta S_E[\,g\,]}{\delta g_{\mu\nu}(x)}
    =M_P^2\,g^{1/2}
    \left(R^{\mu\nu}-\frac12
    g^{\mu\nu}R\right).           \label{2.2}
    \end{eqnarray}

The action is local and manifestly covariant, but it contains
together with the spacetime metric auxiliary structures -- as a
part of boundary conditions it involves at spacetime infinity the
vector field normal to the boundary and the corresponding
extrinsic curvature. As we will now see these structures can be
identically excluded from the action without loosing covariance,
but by the price of locality -- the local action will be
transformed to the manifestly nonlocal form which will serve as a
hint for constructing {\em covariant} long-distance modifications.

Another property of the action (\ref{2.1}) is that it is
explicitly linear in the curvature. However, this linearity is in
essence misleading, because the variational derivative (\ref{2.2})
is also linear in curvature and, therefore, it is at least linear
in metric perturbation on flat-space background $R_{\mu\nu}\sim
h_{\mu\nu}$. Thus the flat-space perturbation theory for the
Einstein action should start with the quadratic order, ${\rm
O}\,[\,h_{\mu\nu}^2\,]\sim {\rm O}\,[\,R_{\mu\nu}^2\,]$. This is a
well known fact from the theory of free massless spin-2 field. Our
goal is to make this $h_{\mu\nu}$-expansion manifestly covariant,
that is to convert it to the covariant (but generally nonlocal)
expansion in powers of the curvature. A systematic way to do this
is to use the technique of covariant perturbation theory of
\cite{beyond,CPTII,basis}. This technique begins with the derivation of
the expression for the metric perturbation in terms of the
curvature and in our context looks as follows.

Expand the Ricci curvature in metric perturbations on flat-space
background
    \begin{eqnarray}
    &&R_{\mu\nu}=-\frac12\,\Box\,h_{\mu\nu}+\frac12\,
    \nabla_\mu\!
    \left(\nabla^\lambda h_{\nu\lambda}
    -\frac12\,\nabla_\nu h\right)\nonumber\\
    &&\qquad\qquad\qquad\qquad\qquad\qquad
    +\frac12\,\nabla_\nu\!
    \left(\nabla^\lambda h_{\mu\lambda}
    -\frac12\,\nabla_\mu h\right)
    +{\rm O}\,[\,h_{\mu\nu}^2\,]         \label{2.3}
    \end{eqnarray}
and solve it by iterations as a nonlocal expansion in powers of
the curvature. This expansion starts with the following terms
    \begin{eqnarray}
    h_{\mu\nu}=-\frac2{\Box}R_{\mu\nu}
    +\nabla_\mu f_\nu+\nabla_\nu f_\mu
    +{\rm O}\,[\,R_{\mu\nu}^2\,].        \label{2.4}
    \end{eqnarray}
Here $1/\Box$ acting on $R_{\mu\nu}$ denotes the action of the
Green's function $G_{\mu\nu}^{\;\;\;\;\alpha\beta}(x,y)$ of the
{\em covariant metric-dependent} d'Alembertian $\Box
\delta^{\alpha\beta}_{\mu\nu}\equiv
g^{\lambda\sigma}\nabla_\lambda\nabla_\sigma
\delta^{\alpha\beta}_{\mu\nu}$ on the space of symmetric
second-rank tensors with natural zero boundary conditions at
infinity
    \begin{eqnarray}
    \frac1{\Box}R_{\mu\nu}(x)\equiv
    \frac{\delta^{\alpha\beta}_{\mu\nu}}{\Box}R_{\alpha\beta}(x)
    =\int dy\,
    G_{\mu\nu}^{\;\;\;\;\alpha\beta}(x,y)\,R_{\alpha\beta}(y),
    \end{eqnarray}
$\Box_x
    G_{\mu\nu}^{\;\;\;\;\alpha\beta}(x,y)=
    \delta^{\alpha\beta}_{\mu\nu}\delta(x,y),\,\,\,\,
    G_{\mu\nu}^{\;\;\;\;\alpha\beta}(x,y)\,\big|_{\,|x|\to\infty}=0$.
In what follows we will not specify the tensor structure of the
Green's functions of $\Box$ implicitly assuming that it is always
determined by the nature of the quantity acted upon by $1/\Box$.

The term $\nabla_\mu f_\nu+\nabla_\nu f_\mu$ in (\ref{2.4})
reflects the gauge ambiguity in the solution of (\ref{2.3}) for
$h_{\mu\nu}$ (originating from the harmonic-gauge terms in the
right-hand side of (\ref{2.3})), but its explicit form is not
important for our purposes here\footnote{The only important
property of this term is that this is a gauge transformation with
some gauge parameter $f_\mu\sim \nabla^\nu h_{\mu\nu}-\nabla_\mu
h/2+{\rm O}\,[h_{\mu\nu}^2]$. Explicit gauge fixing procedure for
the equation (\ref{2.3}) becomes important in higher orders of
curvature expansion and it is presented in much detail in
\cite{beyond,CPTII,basis}.}.

Now restrict ourselves with the approximation quadratic in
$R_{\mu\nu}$ (or equivalently, $h_{\mu\nu}$) and integrate the
variational equation (\ref{2.2}) for $S_E[\,g\,]$. Since the
variational derivative is at least linear in $h_{\mu\nu}$, $\delta
S_E/\delta g_{\mu\nu}\sim h_{\alpha\beta}$, the quadratic part of
the action in view of this equation is given by the integral
    \begin{eqnarray}
    S_E[\,g\,]=\frac12\int dx \,h_{\mu\nu}(x)
    \frac{\delta S_E[\,g\,]}{\delta g_{\mu\nu}(x)}
    +{\rm O}\,[\,R_{\mu\nu}^3\,].           \label{2.5}
    \end{eqnarray}
Substituting (\ref{2.2}) and (\ref{2.4}) and integrating by parts
one finds that the contribution of the gauge parameters $f_\mu$
vanishes in view of the Bianchi identity for the Einstein tensor,
and the final result reads
    \begin{eqnarray}
    S_E[\,g\,]=
    M_P^2\int dx\,g^{1/2}\,\left\{\,
    -\Big(R^{\mu\nu}
    -\frac12\,g^{\mu\nu}R\Big)\,
    \frac1{\Box}R_{\mu\nu}
    +{\rm O}\,[\,R_{\mu\nu}^3\,]\,\right\}.  \label{2.6}
    \end{eqnarray}
This is the covariant {\em nonlocal} form of the {\em local}
Einstein action which was originally observed in our previous
papers on braneworld scenarios with two repulsive branes
\cite{brane,nlbwa}. This nonlocal incarnation of (\ref{2.1})
explicitly features: i) the absence of linear in curvature term
and ii) the absence of auxiliary structures associated with
spacetime infinity. Before we go over to the construction of
long-distance modifications of the theory, let us briefly dwell on
higher-order curvature terms. This, in particular, will clarify
the role played by the Gibbons-Hawking action in the subtraction
of the linear term.

In asymptotically-flat (Euclidean) spacetime with the asymptotic
behavior of the metric
    \begin{eqnarray}
    g_{\mu\nu}=\delta_{\mu\nu}+h_{\mu\nu},\,\,\,\,
    h_{\mu\nu}={\rm
    O}\,\left(\frac1{|x|^{d-2}}\right),\,\,|x|\to\infty,
    \end{eqnarray}
the noncovariant form of the Gibbons-Hawking term in Cartesian
coordinates reads as
    \begin{eqnarray}
    &&S_{\rm GH}[\,g\,]\equiv-2\,M_P^2\int_\infty\!
    d^{d-1}\sigma\,\Big(g^{(d-1)}\Big)^{1/2}\,
    \Big(K-K_0\Big)\nonumber\\
    &&\qquad\qquad\qquad\qquad\qquad\qquad\qquad
    =M_P^2\int\limits_{|x|\to\infty} d\sigma^\mu\,
    \big(\partial^\nu
    h_{\mu\nu}-\partial_\mu h\Big).        \label{2.7}
    \end{eqnarray}
This surface integral can be transformed to the bulk integral of
the integrand $\partial^\mu\big(\partial^\nu
h_{\mu\nu}-\partial_\mu h\Big)$ -- the linear in $h_{\mu\nu}$ part
of the scalar curvature. From the viewpoint of the metric in the
interior of spacetime this is a topological invariant depending
only on the asymptotic behavior $g^\infty_{\mu\nu}=\delta_{mu\nu}+
h_{\mu\nu}(x)\,\big|_{\,|x|\to\infty}$. Similarly to the above
procedure this integral can be covariantly expanded in powers of
the curvature. Up to cubic terms inclusive this expansion
reads\footnote{Validity of this result can be checked either by
the direct $h_{\mu\nu}$-expansion of the right-hand side or by
systematically expanding $h_{\mu\nu}$ on the left hand side as
covariant series in the curvature, starting with (\ref{2.4})
\cite{beyond,CPTII,basis}.}
    \begin{eqnarray}
    \int_\infty\! d\sigma^\mu\,
    \big(\partial^\nu
    h_{\mu\nu}-\partial_\mu h\Big)&=&
    \int dx\,g^{1/2}\left\{\,R
    -\Big(R^{\mu\nu}
    -\frac12\,g^{\mu\nu}R\Big)\,\frac1{\Box}R_{\mu\nu}
    \right.\nonumber\\
       &&+\frac12\,R\left(\frac1{\Box}
    R^{\mu\nu}\right)\frac1{\Box} R_{\mu\nu}
       -R^{\mu\nu}\left(\frac1{\Box}
    R_{\mu\nu}\right)\frac1{\Box} R\nonumber\\
       &&
       +\left(\frac1{\Box} R^{\alpha\beta}\right)
    \left(\nabla_\alpha\frac1{\Box} R\right)
    \nabla_\beta\frac1{\Box} R\nonumber\\
    &&-2\,\left(\nabla^\mu\frac1{\Box} R^{\nu\alpha}\right)
    \left(\nabla_\nu\frac1{\Box}
    R_{\mu\alpha}\right)\frac1{\Box} R \nonumber\\
    &&
    -2\,\left(\frac1{\Box} R^{\mu\nu}\right)
    \left(\nabla_\mu\frac1{\Box}
    R^{\alpha\beta}\right)\nabla_\nu\frac1{\Box}
    R_{\alpha\beta}
    \left.+{\rm O}\,[\,R_{\mu\nu}^4\,]\,\right\}.     \label{2.8}
    \end{eqnarray}
As we see, when substituting to (\ref{2.1}) its linear term
cancels the Ricci scalar part, the quadratic terms reproduce those
of (\ref{2.6}) and the cubic terms recover ${\rm
O}\,[\,R_{\mu\nu}^3\,]$. Obviously, this type of expansion can be
extended to arbitrary order in curvature.

\section{Long-distance modification of the Einstein action}
\hspace{\parindent} Long distance modification of the Einstein
action that would generate (\ref{1.3}) as the left-hand side of
the gravitational equations of motion now can be simply obtained
from the nonlocal form of the Einstein action (\ref{2.6}). It is
just enough to make the following replacement in the quadratic
part of (\ref{2.6})
    \begin{eqnarray}
    \frac1{\Box}\rightarrow
    \frac{1+{\cal F}(L^2\Box\,)}{\Box}.  \label{3.1}
    \end{eqnarray}
Indeed, the subsequent variation of the Ricci tensor, $\delta_g
R_{\mu\nu}=-\frac12\Box\,\delta g_{\mu\nu}+\nabla_\mu
f_\nu+\nabla_\nu f_\mu$, in
    \begin{eqnarray}
    &&\delta_g\int dx\, g^{1/2}
    \Big(R^{\mu\nu}
    -\frac12\,g^{\mu\nu}R\Big)
    \frac{1+{\cal F}(L^2\Box\,)}{\Box}\,
    R_{\mu\nu}=\nonumber\\
    &&\qquad\qquad\qquad\quad 2\int dx\, g^{1/2}
    \Big(R^{\mu\nu}
    -\frac12\,g^{\mu\nu}R\Big)
    \frac{1+{\cal F}(L^2\Box\,)}{\Box}\;
    \delta_g R_{\mu\nu}+\,{\rm O}\,[\,R_{\mu\nu}^2\,]
    \end{eqnarray}
and integration by parts "cancel" the denominator of (\ref{3.1}),
whereas the contribution of gauge parameters $f_\mu$ vanishes, as
above, in view of the Bianchi identity. All commutators of
covariant derivatives with the $\Box$ in the form factor
(\ref{3.1}) give rise to the curvature-squared order which is
beyond our control. This recovers the Einstein tensor term of
(\ref{1.3}) with the needed "nonlocal" Planckian mass
$M_P^2\Big(1+{\cal F}(L^2\Box\,)\Big)$.

The result of the replacement (\ref{3.1}) can be rewritten so that
the contribution of $1$ in the numerator of the new form factor
is again represented in the usual local form of the Einstein
action (\ref{2.1}). Then, the long-distance modification takes the
form of the additional nonlocal term
    \begin{eqnarray}
    S_{NL}[\,g_{\mu\nu}\,]\,
    &=&\,S_E[\,g_{\mu\nu}\,]\nonumber\\
    &&-M_P^2\int dx\,g^{1/2}\,\left\{\,
    \Big(R^{\mu\nu}
    -\frac12\,g^{\mu\nu}R\Big)
    \frac{{\cal F}(L^2\Box\,)}{\Box}\,
    R_{\mu\nu}
    +\,{\rm O}\,[\,R_{\mu\nu}^3\,]\,\right\}.  \label{3.2}
    \end{eqnarray}
This term is not unique though, because it is defined by a given
form factor ${\cal F}(L^2\Box\,)$ only in quadratic order, while
we do not have good principles to fix its higher-order terms thus
far.

This action is manifestly generally covariant. Therefore, its
variational derivative (the left hand side of the modified
Einstein equations) exactly satisfies the Bianchi identity,
    \begin{eqnarray}
    &&\nabla_\mu\frac{\delta S_{NL}[\,g_{\mu\nu}\,]}
    {\delta g_{\mu\nu}(x)}=\nonumber\\
    &&\qquad\qquad
    -M_P^2\,g^{1/2}\nabla_\mu\left[\,\Big(1+{\cal F}(L^2\Box\,)\Big)
    \left(R_{\mu\nu}-\frac12
    g_{\mu\nu}R\right)+{\rm O}\,[\,R_{\mu\nu}^2\,]\,\right]=0,
    \end{eqnarray}
and thus does not suffer from the concerns of \cite{AHDDG}. The
commutator of the covariant derivative with the form factor
$\Big(1+{\cal F}(L^2\Box\,)\Big)$ gives rise to curvature squared
terms and cancels against ${\rm O}\,[\,R_{\mu\nu}^2\,]$.

\section{Discussion: running coupling constants and nonlocality
vs acausality} \hspace{\parindent} One of the main mechanisms for
nonlocalities of the above type is the contribution of graviton
and matter loops to the quantum effective action. In quantum
theory the concept of a nonlocal form factor replacing a coupling
constant is not new. In fact this concept underlies the notion of
the running coupling constants and sheds new light on the
cosmological constant problem also from the viewpoint of the
renormalization theory.

For simplicity, consider QED or Yang-Mills theory in the
quadratic order in gauge field strength $F_{\mu\nu}^2$. The transition
from classical to quantum effective action, $S\to S_{\rm eff}$,
boils down to the replacement of the local invariant by
    \begin{eqnarray}
    \frac1{g^2}\int dx\,F_{\mu\nu}^2\to
    \int dx\,F_{\mu\nu}\,g_{\rm eff}^{-2}(-\Box)\,
    F^{\mu\nu}.
    \end{eqnarray}
Here the effective coupling constant $g_{\rm eff}^{-2}(-\Box)$
is actually a nonlocal form factor playing the role of
${\cal F}(L^2\Box)$ above. It is given in terms of the renormalized
running coupling $g_{R}^2(\mu^2)$ and the beta-function $\beta$
and reads, say in the one-loop approximation, as
    \begin{eqnarray}
    \frac1{g_{\rm eff}^2(-\Box)}=
    \frac1{g_{R}^2(\mu^2)}+
    \beta\,\ln\left(\frac{-\Box}{\mu^2}\right),  \label{4.0}
    \end{eqnarray}
where $\mu^2$ is an auxiliary dimensional parameter. The form
factor $1/g_{\rm eff}^2(-\Box)$ is independent of this parameter
in virtue of the renormalization group equation for
$g_{R}^2(\mu^2)$. Actually, this serves as a basis for the
folklore statement that $\mu^2$ determines the energy scale of the
problem -- quantum effects reduce to the classical effects with
the bare coupling constant $g$ replaced by the running one
$g_{R}(\mu^2)$ at $\mu^2=-\Box$ (this formal substitution in the
Euclidean domain of the form factor (\ref{4.0}) annihilates its
nonlocal logarithmic part and, thus, serves as a qualitative
justification for such an interpretation). Vicy versa, the
knowledge of $g_{R}(\mu^2)$ as a solution of the RG equation
allows one to recover the corresponding nonlocal part of the
effective action.

This concept, though being extremely fruitful in context of gauge
field theory, fails when applied to the gravitational theory in
the sector of the cosmological and Einstein-Hilbert
terms\footnote{When applied to formally renormalizable (albeit
non-unitary) curvature-squared gravitational models
\cite{Tseytlin} or models with generalized renormalization group
for infinite number of charges \cite{BKK}.}. Indeed, naive
replacement of ultralocal cosmological and gravitational coupling
constants by nonlocal form factors
    \begin{eqnarray}
    \int dx\,g^{1/2}\left(\,\Lambda-M_P^2\,R\,\right)\to
    \int dx\,g^{1/2}\left(\,\Lambda(\Box)-M_P^2(\Box)
    \,R\,\right)
    \end{eqnarray}
is meaningless because the action of the covariant d'Alembertian
on the right hand side always picks up its zero mode, and
both form factors reduce to its numerical values in far infrared,
$\Lambda(0)$, $M_P^2(0)$. This happens because the argument of
$\Lambda(\Box)$ has nothing to act upon but 1 (or $g^{1/2}$), and
with $M_P^2(\Box)\,R$ the same happens after integration by parts
\cite{shap}. Therefore, even if one has solutions of 
renormalization group equations 
$\Lambda_{R}(\mu^2)$, $(M_P^2)_{R}(\mu^2)$,
like those obtained in \cite{Tseytlin,BKK}, one cannot automatically
recover the corresponding pieces of effective action or the
corresponding nonlocal correlation functions.

The construction of Sects. 2 and 3 above suggests that the running
coupling constant ``delocalization'' of $M_P^2$ should be done in
the (already nonlocal) representation of the Einstein action
(\ref{2.6}). It is important that its curvature expansion begins
with the quadratic order. Therefore, it explicitly allows one to
insert the nonlocal form factor $M_P^2(\Box)$ {\em between two
curvatures} so that no integration by parts would result in its
degeneration to a trivial constant. It would be interesting to see
how a similar mechanism works for the nonlocal cosmological
``constant'' $\Lambda(\Box)$.\footnote{On dimensional grounds one
should expect that a quadratic action modelling the cosmological
term would read as $\sim\Lambda\int
dx\,g^{1/2}R^{\mu\nu}(1/\Box^2)R_{\mu\nu}$ -- the structure
modifying (\ref{2.6}) by one extra power of $\Box$ in the
denominator. This structure (also suggested in \cite{shap}
and discussed within the renormalization group theory) appears 
in two-brane models \cite{nlbwa} and as a covariant completion 
of the mass term in
models of massive gravitons \cite{completion} and numerous
discussions of the van Damm-Veltman-Zakharov discontinuity
\cite{vDVZ}.} Nontrivial mechanisms of its generation due to
infrared asymptotics of the effective action, or late-time
asymptotics of the corresponding heat kernel, are discussed in
\cite{nneag}.

Nonlocalities of the type (\ref{3.2}) also arise in a certain
class of braneworld models \cite{GRS,DGP,ParSol}. They cannot
appear in models of the Randall-Sundrum type with strictly
localized zero modes, because in these models nontrivial form
factors basically arise in the transverse-traceless sector of the
action (as kernels of nonlocal quadratic forms in {\em Weyl}
tensor \cite{nlbwa}). In contrast to these models, the nonlocal
part of (\ref{3.2}) is not quadratic in the Weyl tensor, $\int
dx\,g^{1/2}W^2_{\mu\nu\alpha\beta}\sim\int
dx\,g^{1/2}(R_{\mu\nu}^2-\frac13 R^2)$ (with the insertion of a
nonlocal form factor between the curvatures). Rather, (\ref{3.2})
includes the structure $\int dx\,g^{1/2}(R_{\mu\nu}^2-\frac12
R^2)$ which contains the {\em conformal} sector. It is this sector
which is responsible for the potential resolution of the
cosmological constant problem. It becomes dynamical in models with
metastable graviton like the Gregory-Rubakov-Sibiryakov model
\cite{GRS} or Dvali-Gabadaze-Porrati model (DGP) \cite{DGP}. In
particular, for the (4+1)-dimensional DGP model the (Euclidean)
form factor ${\cal F}(L^2\Box)$ is singular at $\Box\to 0$ and has
the form \cite{DGP,DefDG,DGZ}
    \begin{eqnarray}
    M_P^2\,{\cal F}(L^2\Box)=
    \frac{M^3}{\sqrt{-\Box}},\,\,\,\,\,
    M^3=\frac{M_P^2}L,              \label{4.1}
    \end{eqnarray}
where $M\sim 10^{-21}M_P\sim 100$ MeV is a mass scale of the bulk
gravity as opposed to the Planckian scale of the Einstein term on
the brane $M_P\sim 10^{19}$ GeV.

Both form factors (\ref{4.0}) and (\ref{4.1}) are unambiguously
defined only in the Euclidean space where the d'Alembertian $\Box$
is negative semi-definite. This raises the problem of their
continuation to the physical spacetime where the issues of
causality and unitarity become important. The principles of this
continuation depend on the physical origin of nonlocality in
${\cal F}(L^2\Box)$. Depending on whether it has a quantum nature
like in (\ref{4.0}) or brane induced nature like in (\ref{4.1})
these principles can range from the usual Wick rotation to such
currently developing paradigms as holographic dS/CFT-conjecture
\cite{dS/CFT}, the concept of time as a holographically generated
dimension, etc. \cite{AHDDG}. Let us first discuss nonlocal form
factors of quantum nature.

Scattering problems for in-out matrix elements of quantum field
$\hat\varphi$, $\big<\,out\,|\,\hat\varphi\,|\,in\,\big>$, in
spacetime with asymptotically-flat past and future imply a usual
Wick rotation. The expectation-value problem or the problem for
in-in mean value of the quantum field,
$\phi=\big<\,in\,|\,\hat\varphi\,|\,in\,\big>$, is more
complicated and incorporates the Schwinger-Keldysh diagrammatic
technique \cite{SchKel}. In this technique the effective equations
for $\phi$ cannot be obtained as variational derivatives of some
action functional\footnote{In the general case they can be
obtained by varying a special two-field functional with respect to
one field, the both fields subsequently being set coincident and
equal to the mean field in question \cite{SchKel}.}. So for this
problem the action as a source of effective equations does not
exist at all. However, there exists a special case of the quantum
initial data in the form of the {\em Poincare-invariant in-vacuum}
in asymptotic past, $|\,in\,\big>=|\,in,vac\,\big>$. Effective
equations for $\phi$ in this vacuum can be obtained by the
following procedure \cite{beyond}. Calculate the {\em Euclidean}
effective action in asymptotically-flat spacetime, take its
variational derivative containing the nonlocal form factors which
are uniquely specified by zero boundary conditions at Euclidean
infinity. Then formally go over to the Lorentzian spacetime
signature with the {\em retardation} prescription for all nonlocal
form factors. These retarded boundary conditions uniquely specify
the nonlocal effective equations and guarantee their causality.
This procedure was proven in \cite{beyond} and also put forward in
a recent paper \cite{Woodard} as the basis of the covariant
nonlocal model of MOND theory \cite{MOND}.

As we see, this procedure justifies the Euclidean setup used above
and suggests straightforward appplications in the
expectation-value problem of the above type. Interestingly, in
this setting no contradiction arises between the nonlocal nature
of the {\em Euclidean} action and {\em causal} nature of nonlocal
equations of motion in Lorentzian spacetime. In this respect the
situation is essentially different from the assumptions of
\cite{AHDDG} where acausality of equations of motion is
necessarily attributed to the nonlocal action. In fact, this
property requires a detailed analysis of why only the phenomena
slow at the cosmological scale $L$ turn out to be acausal, while
the phenomena generated by ``small'' sources $(\ll L)$ are
essentially causal. No such assumptions are needed in effective
equations for expectation values which are fundamentally causal
despite their nonlocality. These equations have interesting
applications in quantum gravitational context and, in particular,
show the phenomenon of the cosmological acceleration due to
infrared back-reaction mechanisms \cite{W1}.

The situation with brane induced nonlocalities and their causality
status is more questionable and conceptually open. For example,
the branch point of the square root in the nonlocal form factor
(\ref{4.1}) is apparently related to different branches of
cosmological solutions including the scenario of cosmological
acceleration \cite{DefDG}. Therefore, in contrast to tentative
models of \cite{AHDDG} with finite ${\cal F}(0)\gg 1$, which only
interpolate between two Einstein theories with different
gravitational constants $G_{LD}\sim G_P/{{\cal F}(0)}\ll G_P$, the
DGP model is anticipated to suggest the mechanism of the
cosmological acceleration. This implies the replacement of the
asymptotically-flat spacetime by the asymptotically-deSitter one.
For small values of asymptotic curvature (as is the case of the
observable horizon scale $H_0^2/M_P^2\sim 10^{-120}$) the
curvature expansion used above seems plausible, although the
effect of the asymptotic curvature might be in essence
nonperturbative. Therefore, the above construction might have to
be modified. In particular, the Gibbons-Hawking term should be
replaced by its asymptotically-deSitter analogue and the expansion
in powers of the curvature should be replaced by the expansion in
powers of its deviation from the asymptotic value
$R_{\mu\nu}-\frac1d g_{\mu\nu}R_\infty$. This would introduce in
the formalism as a free parameter the value of the curvature in
far future, $R_\infty\sim H_0^2$, reflecting the measure of
acausality in the model. It might be related to the CFT central
charge, $c_{UV}=M_P^2/R_\infty\sim 10^{120}$ \cite{AHDDG} -- the
number of holographic degrees of freedom in dS/CFT conjecture. The
resulting modifications in the above construction are currently
under study and will be presented elsewhere.

\section*{Acknowledgements}

The author would like to thank V.Mukhanov, S.Solodukhin and
R.Woodard for helpful stimulating discussions and is grateful to
G.Gabadadze for useful comments on the first draft of this paper.
He is also grateful for hospitality of the Physics Department of
LMU, Munich, where a part of this work has been done under the
grant SFB375. This work was also supported by the Russian
Foundation for Basic Research under the grant No 02-02-17054 and
the LSS grant No 1578.2003.2.

\end{document}